# A computer code for forward calculation and inversion of the H/V spectral ratio under the diffuse field assumption

Antonio García-Jerez[a,b], José Piña-Flores[c], Francisco J. Sánchez-Sesma[c], Francisco Luzón[a], and Mathieu Perton[c]

**Abstract**

During a quarter of a century, the main characteristics of the horizontal-to-vertical spectral ratio of ambient noise HVSRN have been extensively used for site effect assessment. In spite of the uncertainties about the optimum theoretical model to describe these observations, over the last decade several schemes for inversion of the full HVSRN curve for near surface surveying have been developed.

In this work, a computer code for forward calculation of H/V spectra based on the diffuse field assumption (DFA) is presented and tested. It takes advantage of the recently stated connection between the HVSRN and the elastodynamic Green's function which arises from the ambient noise interferometry theory.

[a] Departamento de Química y Física, Universidad de Almería. 04120 Almería, Spain.   Emails: agarcia-jerez@ual.es, fluzon@ual.es

[b] Instituto Andaluz de Geofísica. Universidad de Granada. C/ Profesor Clavera, 12. 18071 Granada, Spain.

[c] Instituto de Ingeniería, Universidad Nacional Autónoma de México, CU, Coyoacán, 04510 D.F., Mexico. Emails: ead2009@hotmail.com, sesma@unam.mx, mathieu.perton@gmail.com

The algorithm allows for (1) a natural calculation of the Green's functions imaginary parts by using suitable contour integrals in the complex wavenumber plane, and (2) separate calculation of the contributions of Rayleigh, Love, P-SV and SH waves as well. The stability of the algorithm at high frequencies is preserved by means of an adaptation of the Wang's orthonormalization method to the calculation of dispersion curves, surface-waves medium responses and contributions of body waves.

This code has been combined with a variety of inversion methods to make up a powerful tool for passive seismic surveying.

**1. Introduction**

Since the early work of Kanai et al. (1954), many efforts have been devoted to the observation and interpretation of the seismic ambient noise (microtremor), consisting of background vibrations due to natural phenomena of atmospheric, oceanic, seismic and volcanic origins as well as human activities like traffic and industry. The possibility of using this ubiquitous natural illumination for exploration of ground structures with seismic arrays was soon recognized by Aki (1957).

Later, the capabilities of single-station measurements of this wavefield were enhanced by Nakamura (1989), who proposed that the ratio between the spectra of horizontal and vertical components of ambient noise allowed for identification of resonance peaks and for estimation of seismic amplifications of soils. After these early developments, several sophisticated tools have been introduced to take full advantage of these spectral ratios for seismic exploration, which are based on diverse and in some cases opposite theoretical approaches.

Fäh et al. (2003) and Wathelet (2005) studied the inversion of the Rayleigh wave ellipticity, sometimes considered as a rough proxy of the horizontal-to-vertical spectral ratio of ambient noise (HVSRN), using genetic algorithms and the neighborhood method as inversion procedures. Their method was not intended to be applied using energy ratios of raw microtremor records, because they consist of a complicated mixture of Rayleigh, Love and body waves, but is better suited when used together with more advanced processing aimed at extracting the ellipticity (*e.g.* Poggi et al., 2012).

Arai and Tokimatsu (2004) introduced a method for inversion of the HVSRN accounting for surface waves generated by a continuum of uncorrelated shallow sources located far enough from the receiver. Their approximate expressions allow for a quick and suitable estimation of the power ratio in those models and frequency bands where surface waves are the dominant contribution. Related full-wavefield versions have been developed by Lachet and Bard (1994) and Lunedei and Albarello (2009). To the best of the authors' knowledge, these two methods have not been incorporated into ground inversion tools. It is probably due to the intensive computing requirements.

On the other hand, a code for inversion of layered ground structures from horizontal-to-vertical spectral ratios of microtremor was published by Herak (2008). The algorithm models the HVSRN as the ratio between the ground responses for vertically incident S and P waves. In this approach, the anelastic attenuation plays a major role in order to make the H/V decay at high frequencies as observed in most experimental conditions. This can be regarded as an oversimplified scheme taking into account the predominance of surface waves found in broad frequency bands of the microtremor spectra.

Sánchez-Sesma et al. (2011) introduced an innovative method inspired in the possibility of retrieving the 3D elastodynamic Green's tensor between two stations embedded in an elastic medium from the average time-domain cross-correlation of their ambient noise records (ambient noise interferometry). Some applications of this method have been developed by Salinas et al. (2014), Kawase et al. (2015), Lontsi et al. (2015), Rivet et al. (2015), Spica et al. (2015), among others. The theoretical foundations of this general theory were developed in several research works (e.g. Snieder, 2004; Wapenaar, 2004; Sánchez-Sesma and Campillo, 2006) and confirmed in experiments with microtremors by Shapiro and Campillo (2004). The equations used here for modelling the HVSRN appear naturally in the particular case of the interstation distance tending to zero. Sánchez-Sesma et al. (2011) formulated a first algorithm for forward calculation of the HVSRN under this approach based on the discrete wavenumber integration and the matrix method described by Knopoff (1964). Equivalent results can be obtained using a generalized view of the modal equipartition proposed by Margerin (2009).

A more complete review of these theories appeared recently in Lunedei and Malischewsky (2015). The authors made a thorough description of the various theoretical models explaining the HVSRN that developed in the last three decades, including the diffuse field assumption, and compared the corresponding theoretical curves for a set of test models.

In this communication we present a faster code which allows separate computation of the contributions of SH, P-SV, Rayleigh and Love waves together with algorithms for inversion of ground structures. The software is available at the hv-inv project website http://www.ual.es/GruposInv/hv-inv/ as FORTRAN 90 and Matlab codes. In a companion paper (Piña-Flores et al., 2016) we explore the properties of HVSRs as well as the non-uniqueness problem.

**2. Forward calculation of the H/V spectral ratio**

The algorithms described here are based on the representation of the ambient noise wavefield as a 3D-diffuse vector field established within the medium. Then, the directional power spectral densities (PSD) of motion $P_m(\mathbf{x};\omega)$ along any Cartesian axis $m$ at an arbitrary point $\mathbf{x}$ and circular frequency $\omega$ are found to be proportional to the imaginary part of the diagonal components of the Green's tensor at $\mathbf{x}$ when both source and receiver coincide, $\text{Im}[G_{mm}(\mathbf{x};\mathbf{x};\omega)]$ (no sum, Sánchez-Sesma et al., 2008). When the ratio between two distinct directional PSDs is considered, the proportionality factor cancels out. Then, the ratio between the horizontal (sum of the terms associated with directions 1 and 2, which in fact are equal) and the vertical directional PSDs (term associated with direction 3), that is, the usual H/V spectral amplitude ratio at point $\mathbf{x}$, corresponds to:

$$[H/V](\mathbf{x};\omega) \equiv \sqrt{\frac{2P_1(\mathbf{x};\omega)}{P_3(\mathbf{x};\omega)}} = \sqrt{\frac{2\,\mathrm{Im}[G_{11}(\mathbf{x};\mathbf{x};\omega)]}{\mathrm{Im}[G_{33}(\mathbf{x};\mathbf{x};\omega)]}}. \quad (1)$$

The middle expression arises from the noise auto-correlations; instead, the right hand side may be calculated theoretically under some properties and geometrical assumptions. In this work, the configuration considered is a horizontally layered structure on top of a half-space (Fig. 1). The top surface is free and the layers correspond to homogeneous, elastic and isotropic media with plane interfaces. Quantities $\alpha_j$, $\beta_j$, $\rho_j$, $\mu_j = \rho_j \beta_j^2$ and $h_j$ stand for P- and S-wave velocities, mass density, shear modulus and thickness of the $j$-th layer, respectively. Under these assumptions, a convenient way to evaluate the Green's function components in Eq. (1), on the basis of frequency-horizontal wavenumber $(\omega, k)$ representation of the wavefield, consists in using the contour integration method in the complex wavenumber plane. That method provides a rigorous way for dealing with the singularities of the original integral expressions defined on the positive $k$ values (e.g. Harkrider, 1964).1

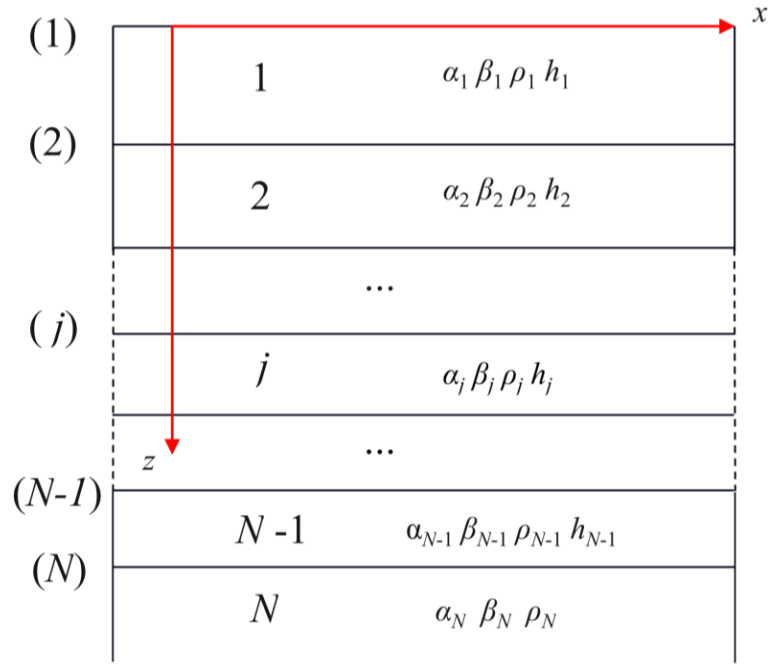

**Figure 1.** Plane layered structure made of isotropic elastic layers. $\alpha_j$, $\beta_j$, $\rho_j$ and $h_j$ stand for P- and S-wave velocities, mass density and thickness of the $j$-th layer. The numbering of layer and interfaces is shown.

To use the contour integration, the integrals are first extended to the whole real $k$ axis using integrand symmetries. Then, the contour shown in Fig. 2 (Tokimatsu and Tamura, 1995; García-Jerez et al., 2013) can be employed to isolate the contributions to the Green's functions of the body and any of the surface waves modes. The body wave contributions are calculated from the integral along the branch-cuts, whereas the surface waves modes contributions are proportional to the residues at the poles located on the positive real $k$ axis.

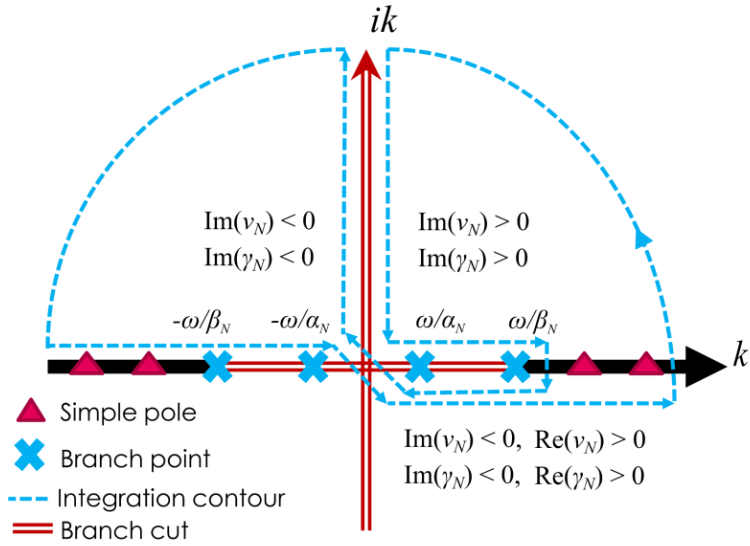

**Figure 2.** Contour used for complex-plane integration, from Tokimatsu and Tamura (1995).

For coincident source and receiver, the sum of the branch-cut integrals along both sides of the positive imaginary axis contributes to the (diverging) real part of the Green's functions, whereas the integrals along the segments of the branch-cuts on the real axis contribute to the imaginary part. Since Eq. (1) involves $\mathrm{Im}[G_{jj}(\mathbf{x};\mathbf{x};\omega)]$ only, the required numerical integration, corresponding to body waves, is constrained to the interval $[0, \omega/\beta_N]$. This may significantly speed up the computations comparing with the direct application of the conventional discrete wavenumber method, even though the actual improvement depends on the computational effort required to find the poles (dispersion curves) which varies from model to model.

The contributions of the residues at the poles and of the branch-cut integrals to the imaginary part of the Green's function at source are:

$$\mathrm{Im}[G_{33}(0;0;\omega)] = -\frac{1}{2}\sum_{m\in\mathrm{RAYLEIGH}} A_{Rm} + \frac{1}{2\pi}\int_0^{\omega/\beta_N} \mathrm{Re}\left[ik\, r_2(k,\omega)\right]_{4^{th}} dk, \qquad (2)$$

$$\mathrm{Im}[G_{11}^{PSV}(0;0;\omega)] = \mathrm{Im}[G_{22}^{PSV}(0;0;\omega)] = -\frac{1}{4}\sum_{m\in\mathrm{RAYLEIGH}} A_{Rm}\chi_m^2$$

$$+ \frac{1}{4\pi}\int_0^{\omega/\beta_N} \mathrm{Re}\left[ik\, r_1(k,\omega)\right]_{4^{th}} dk, \qquad (3)$$

$$\mathrm{Im}[G_{11}^{SH}(0;0;\omega)] = \mathrm{Im}[G_{22}^{SH}(0;0;\omega)] = -\frac{1}{4}\sum_{m\in\mathrm{LOVE}} A_{Lm} + \frac{1}{4\pi}\int_0^{\omega/\beta_N} \mathrm{Re}\left[ik\, l_1(\omega,k)\right]_{4^{th}} dk, \qquad (4)$$

where the real quantities $A_{Rm}$ and $A_{Lm}$ stand for the medium responses for the *m*-th Rayleigh and Love modes, respectively. The expressions for the contributions of surface waves result from the limit $r \to 0$ of the general formulae derived by Harkrider (1964). The integral terms may be interpreted as the motion at surface due to wave systems for which S waves spread through the halfspace from vertically to horizontal incidence. The integrands are evaluated on the fourth-quadrant side of the real *k* axis. In most cases, but not always, the integrands in Eqs. (2-4) are free from poles or sharp peaks so that the computational effort for numerical integration is moderate. Some exceptions to this rule, due to the presence of leaky-mode branches in the proximity of the real *k* axis have been studied by García-Jerez and Sánchez-Sesma (2015).

On the other hand, it is well-known that the Thomson-Haskell propagator-matrix method presents a weakness related with numerical overflow and loss of precision at high frequencies (e.g. Buchen and Ben-Hador, 1996). Algorithms for stabilization based on Dunkin (1965) have been used in various popular codes for computation of surface waves and Green's functions (e.g. Herrmann and Ammon, 2003; Wathelet, 2005). In the program described here, this question is addressed with an adaptation of the Wang's (1999) orthonormalization method, which preserves the 4 x 4 character of the propagators in the cases of P-SV and Rayleigh waves. The rest of this section is devoted to describe in detail the computations of the elements of Eqs. (2 to 4).

## 2.1. Computation of the surface waves contributions

### 2.1.1. Rayleigh waves dispersion curves

The calculation of these curves is based on the Thomson-Haskell propagator-matrix method following the formulation in Aki and Richards (2002) for propagation of plane Rayleigh waves in the x-z plane (z axis directed down).

First, the quantities $\gamma_j = \sqrt{k^2 - \omega^2/\alpha_j^2}$ and $\nu_j = \sqrt{k^2 - \omega^2/\beta_j^2}$, and the matrix $L_j^{PSV}$

$$L_j^{PSV} = \omega^{-1}\begin{pmatrix} \alpha_j k & \beta_j \nu_j & \alpha_j k & \beta_j \nu_j \\ \alpha_j \gamma_j & \beta_j k & -\alpha_j \gamma_j & -\beta_j k \\ -2\alpha_j \mu_j k \gamma_j & -\beta_j \mu_j (k^2 + \nu_j^2) & 2\alpha_j \mu_j k \gamma_j & \beta_j \mu_j (k^2 + \nu_j^2) \\ -\alpha_j \mu_j (k^2 + \nu_j^2) & -2\beta_j \mu_j k \nu_j & -\alpha_j \mu_j (k^2 + \nu_j^2) & -2\beta_j \mu_j k \nu_j \end{pmatrix} \quad (5)$$

are defined for each layer. After multiplying the columns of $L_j^{PSV}$ by $e^{-\gamma_j(z-z_j)}$, $e^{-\nu_j(z-z_j)}$, $e^{\gamma_j(z-z_j)}$ and $e^{\nu_j(z-z_j)}$, respectively and by the horizontal propagation factor $e^{i(kx-\omega t)}$, they form a basis for the waves in the $j$-th layer displacement-stress space $(r_1, r_2, r_3, r_4)^T = (u_x, -iu_z, \sigma_{zx}, -i\sigma_{zz})^T$ (Aki and Richards, 2002, Eq. 7.55).

Since the phase velocity of surface waves cannot be higher than $\beta_N$, the quantities $\gamma_N$ and $\nu_N$ will be real. In that case, the waves within the halfspace are inhomogeneous. Taking into account these dependences on z, we first define the 4 x 2 coefficients matrix in the halfspace as

$$D_N = \begin{pmatrix} 1 & 0 & 0 & 0 \\ 0 & 1 & 0 & 0 \end{pmatrix}^T, \tag{6}$$

so that the product $Y_N^{PSV} = L_N^{PSV} D_N$ selects the two first columns of $L_N^{PSV}$ which form a basis of the subspace of the vectors, fulfilling the radiation conditions that imply the absence of waves with unbounded amplitude at infinite depth.

To propagate this basis of the subspace upwards, fulfilling the conditions of continuity of stress and displacements across the *N*-th interface, $Y_N$ is first written in terms of the full basis associated to the (*N*-1)-th layer by means of the coefficient matrix

$$D_{N-1} = \left[L_{N-1}^{PSV}\right]^{-1} Y_N^{PSV}. \tag{7}$$

In principle, the next step would be to obtain a basis $Y_{N-1}^{PSV}$ at the top of the (*N*-1)-th layer as the two columns of $L_{N-1}^{PSV} E_{N-1}^{-1} D_{N-1}$, where

$$E_j^{-1} \equiv E_j(-h_j) = \begin{pmatrix} e^{\gamma_j h_j} & 0 & 0 & 0 \\ 0 & e^{\nu_j h_j} & 0 & 0 \\ 0 & 0 & e^{-\gamma_j h_j} & 0 \\ 0 & 0 & 0 & e^{-\nu_j h_j} \end{pmatrix}. \tag{8}$$

Nevertheless, when $\gamma_j$ and $\nu_j$ are (positive) reals, this multiplication involves operations between the two different increasing exponentials which can cause loss of precision for a thick enough layer. Following Wang (1999), this difficulty can be avoided by means of a transformation of $D_{N-1}$ by a 2x2 orthonormalization matrix defined as

$$Q = f(D) = \begin{pmatrix} D_{22} & -D_{12} \\ -D_{21} & D_{11} \end{pmatrix} / \sqrt{(|D_{11}|^2 + |D_{21}|^2)(|D_{12}|^2 + |D_{22}|^2)} . \qquad (9)$$

Thus, $D'_{N-1} = D_{N-1} Q_{N-1}$ has zeroes at the elements 12 and 21, allowing the separation of the exponentials $e^{-\gamma_{N-1}(z-z_N)}$ and $e^{-\nu_{N-1}(z-z_N)}$ along the (N-1)-th layer in the columns of $L_{N-1}^{PSV} E_{N-1}(z-z_N) D'_{N-1}$. Consequently, the alternative (stabilized) basis at $z_{N-1}$ is:

$$Y_{N-1}^{PSV} = L_{N-1}^{PSV} E_{N-1}(-h_{N-1}) D'_{N-1} = L_{N-1}^{PSV} E_{N-1}^{-1} \{ [L_{N-1}^{PSV}]^{-1} Y_N^{PSV} Q_{N-1} \} . \qquad (10)$$

The procedure is then iterated up to the free surface, so that the result is formally equivalent to

$$Y_1^{PSV} = \{ L_1^{PSV} E_1^{-1} [L_1^{PSV}]^{-1} \} \{ L_2^{PSV} E_2^{-1} [L_2^{PSV}]^{-1} \} ... \{ L_{N-1}^{PSV} E_{N-1}^{-1} [L_{N-1}^{PSV}]^{-1} \} L_N^{PSV} D_N Q_{N-1,acc} \qquad (11)$$

where $Q_{j,acc} = Q_j Q_{j-1} ... Q_2 Q_1$ stand for the accumulated transformation from the j-th layer to the surface ($Q_{N-1,acc}$ includes all the layers).

At this point, the dispersion curves could be obtained in terms of $k$ and $\omega$ from the compatibility condition $(Y_1^{PSV})_{33} (Y_1^{PSV})_{44} - (Y_1^{PSV})_{34} (Y_1^{PSV})_{43} = 0$, which guarantees the existence of a non-trivial linear combination of the columns of $Y_1^{PSV}$ fulfilling the conditions of zero stress components $\sigma_{zx}(z_1) = 0$ and $\sigma_{zz}(z_1) = 0$. Nevertheless, due to the stabilization scheme, this relation involves complex quantities $(Y_1^{PSV})_{jk}$ that would imply a more laborious root-finding task. In order to deal with real quantities, it is preferable to state an equivalent equation by using the corresponding elements of $Y_1^{PSV} Q_{N-1,acc}^{-1}$. Moreover, since any positive factor preserves the sign of the determinant and $\text{sign}[\det(Q_{N-1,acc}^{-1})] = \text{sign}[\det(Q_{N-1,acc}^*)] / |\det(Q_{N-1,acc})|^2 ] = \text{sign}[\det(Q_{N-1,acc}^*)]$, we in practice solve the equation:

$$\det\left\{\begin{bmatrix}(Y_1^{PSV})_{31} & (Y_1^{PSV})_{32} \\ (Y_1^{PSV})_{41} & (Y_1^{PSV})_{42}\end{bmatrix} Q_{N-1,acc}^*\right\} = 0. \qquad (12)$$

The signs of this expression are evaluated on the frequency - velocity plane and shown in Fig. 3a for an example model. The improvements in stability due to the orthonormalization scheme are evident in the short-wavelength region.

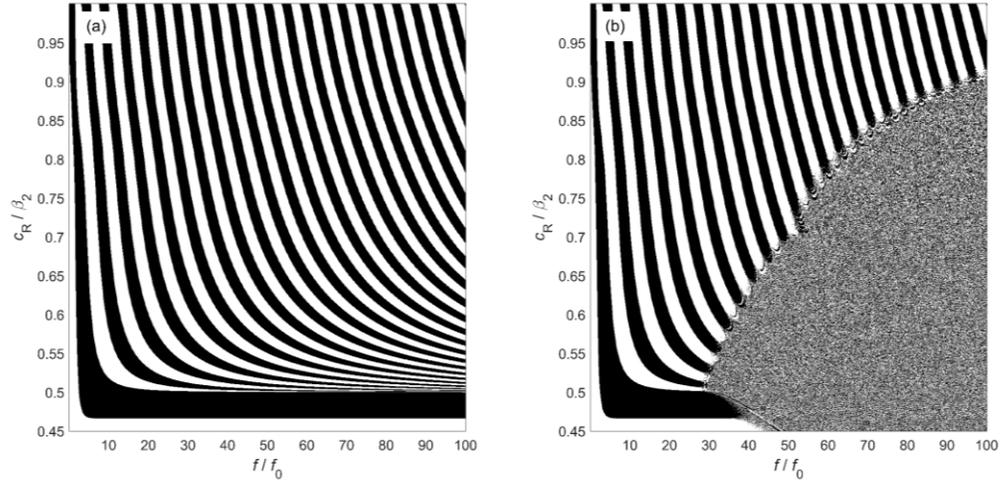

**Figure 3.** (a) Sign of the determinant (frequency equation) used in this work for location of Rayleigh-waves phase velocities (Eq. 12) for an example model. Black color stands for negative and white for positive. (b) Result obtained without introducing the orthonormalization matrices $Q$. The model consists in a single layer over a stiffer halfspace with a contrast of 2 in $\beta$, $\alpha = 2\beta$, and no density contrast. A frequency $f_0 = \beta_1/(4h)$ is used for normalization.

### 2.1.2. Love waves dispersion curves

The scheme followed for calculation of Love-waves dispersion curves parallels the algorithm used for Rayleigh-waves, even though the way for avoiding numerical overflow is simpler.

In this case, the out-of-plane displacement-stress vector $(l_1, l_2)^T = (u_y, \sigma_{yz})^T$ in the $j$-th layer can be written as the columns of

$$L_j^{SH} = \begin{pmatrix} 1 & 1 \\ -v_j \mu_j & v_j \mu_j \end{pmatrix}, \tag{13}$$

with respective dependences on $e^{i(kx-\omega t)-v_j(z-z_j)}$ and $e^{i(kx-\omega t)+v_j(z-z_j)}$. The first one is itself a basis for the harmonic waves in the halfspace fulfilling the boundary conditions, so that $D_N^{SH} = (1 \ 0)^T$. Then, the iterative procedure for upwards propagation described above is applied. In this case, once the corresponding 2x1 coefficient vector for the $j$-th layer basis is calculated as $D_j^{SH} = [L_j^{SH}]^{-1} Y_{j+1}^{SH}$, it is subsequently normalized dividing by $\sqrt{|(D_j^{SH})_1|^2 + |(D_j^{SH})_2|^2}$. Finally, the frequency equation arises as the zero-stress condition at the free surface: $\sigma_{yz}(z = z_1) = (Y_1^{SH})_2 = 0$.

### 2.1.3. Medium response for surface waves

Instead of doing a direct evaluation of the residues by using L'Hôpital's rule (see expressions in Harkrider, 1964) which requires numerical derivatives, the computations of the medium responses for surface waves modes $A_{Rm}$, $A_{Lm}$ (Eqs. 3-4) are carried out by analytical evaluation of energy integrals along the modal shapes at pairs $(\omega, c = c_{Rm}(\omega))$ determined from the dispersion curves (Harkrider and Anderson, 1966).
According to Section 2.1.1, the modal shapes for Rayleigh waves can be calculated in the $j$-th layer as

$$y^R(z) = L_j^{PSV} E_j(z - z_{j+1}) D_j' Q_{acc, j-1} \begin{pmatrix} a \\ 1 \end{pmatrix}, \text{ with } z_j \leq z \leq z_{j+1} \text{ and } a = \frac{-(Y_1^{PSV})_{32}}{(Y_1^{PSV})_{31}}. \tag{14}$$

The modal index $m$ and the layer index $j$ in $y^R(z)$ are dropped off for the sake of simplicity. The multiplication by $(a \ 1)^T$ in Eq. (14) generates a linear combination of the basis $Y_1^{PSV}$ that fulfils the zero-stress condition at surface, whereas the multiplication by $Q_{acc, j-1}$ transforms $D_j{'}$ into the surface basis $Y_1^{PSV}$ (since no transformation is required for $z$ within the first layer, $Q_{acc,0}$ can be defined as the identity matrix). The modal shapes of Love waves can be computed using an analogous (but simpler) scheme.

The calculation described above has been checked in many test models, showing accuracy and stability. Figures 4d-e show the perfect matching with the results of a global matrix method for the model defined by Denolle et al. (2012) in similar tests (Model 1 in Table 1). An example of the instability of the original propagator matrix method as frequency grows is shown in Figs. 4a-c (black lines).

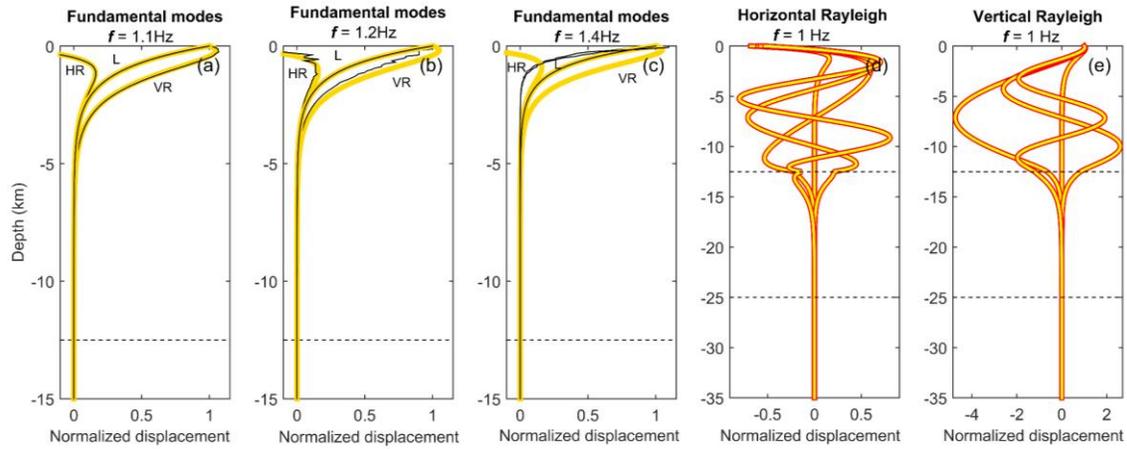

**Figure 4.** (a-c) Comparison between the shapes of the fundamental mode for model 1 (Table 1), computed by means of the algorithm described in this work (yellow line) and with a non-stabilized scheme (black line) based on Aki and Richards (2002). HR, VR and L stand for Horizontal component of Rayleigh wave, Vertical component of Rayleigh wave and Love wave displacements, respectively. (d-e) Shapes of the fundamental and the first three higher modes of Rayleigh waves compared with reference curves (red lines) obtained with a global matrix method (Kausel, 2015, personal communication).

This scheme is subsequently used to compute the integrals (Ben-Menahem and Singh 2000, Eqs. 3.139) for Rayleigh wave modes:

$$I_0^R = \sum_{j=1}^{N} \rho_j \int_{z_j}^{z_{j+1}} \{|u_x(z)|^2 + |u_y(z)|^2\} dz, \tag{15}$$

$$I_1^R = \sum_{j=1}^{N} \mu_j \int_{z_j}^{z_{j+1}} \{\frac{\alpha_j^2}{\beta_j^2} |u_x(z)|^2 + |u_z(z)|^2\} dz \tag{16}$$

and

$$I_3^R = \sum_{j=1}^{N} \mu_j \int_{z_j}^{z_{j+1}} \{\frac{\alpha_j^2}{\beta_j^2} \left|\frac{du_z(z)}{dz}\right|^2 + \left|\frac{du_x(z)}{dz}\right|^2\} dz, \tag{17}$$

where $u_x(z) = r_1(z)$ and $u_z(z) = i r_2(z)$ refer to the displacement components of the modal vector $y^R = (r_1, r_2, r_3, r_4)^T$.

Then, the group velocity is calculated from these integrals and from the phase velocity $c_R$ as

$$v_{gR} = \frac{c_R^2 I_0^R - c_R^2 I_3^R / \omega^2 + I_1^R}{2 I_0^R c_R}, \tag{18}$$

which can be worked out from Ben-Menahem and Singh (2000, Eqs. 3.139, 3.140, 3.141 and 5.128). Note that numerical derivatives of the phase-velocity dispersion curves are avoided in this way. Then, the medium response is subsequently computed as

$$A_R = \frac{|u_z(z_1)|^2}{2 v_{gR} c_R I_0^R}. \tag{19}$$

In the case of Love waves, the relevant energy integrals to be evaluated are

$$I_0^L = \sum_{j=1}^{N} \rho_j \int_{z_j}^{z_{j+1}} |u_y(z)|^2 dz, \tag{20}$$

and

$$I_1^L = \sum_{j=1}^{N} \mu_j \int_{z_j}^{z_{j+1}} |u_y(z)|^2 dz \qquad (21)$$

whereas the medium response can be worked out as:

$$A_L = \frac{|u_y(z_1)|^2}{2 I_1^L} \qquad (22)$$

(Ben-Menahem and Singh 2000, Eqs. 3.128, 3.129 and 5.126). For joint inversion purposes, the group velocity of Love waves can also be calculated as

$$v_{gL} = \frac{I_1^L}{I_0^L c_L}. \qquad (23)$$

Since calculations of medium responses are independent for each frequency and mode, the computation can be easily parallelized to improve performance in shared memory multiprocessing systems. In this manner, each thread deals with a subset of frequencies. This feature has been implemented by means of OpenMP constructs. This also applies to the body wave contributions.

### 2.2. Computation of body waves contributions

The last step is to evaluate the integral terms in Eqs. (2 to 4), which address the effects of the diffuse P-SV and SH body wavefield, respectively. Thus, for any fixed $\omega$ and for both P-SV and the SH contributions, the wavenumber $k$ is uniformly varied from 0 to $\omega/\beta_N$, corresponding to outgoing S waves spreading from vertically to horizontally in the halfspace. The P waves in the halfspace can either be of homogeneous or inhomogeneous type.

For each couple $(\omega, k)$, the boundary conditions at the halfspace and the continuity of stress and displacements through the structure are stated, except at the *source* (evaluation point) located at surface. In particular, respective bases for the displacement-stress vectors $(r_1, r_2, r_3, r_4)^T$ and $(l_1, l_2)^T$ at $z = 0$ are generated by following procedure in Eqs. (6 to 11) for P-SV waves, and that in section 2.1.2 for SH waves.

The constant factors in Eqs. (2-4) have been fitted to represent discontinuities $\delta r_3$, $\delta r_4$ and $\delta l_2$ corresponding to unitary forces, that is, $\delta r_4 = F_z = 1$, and $\delta r_3 = \delta l_2 = F_x = 1$. These factors can be derived from the more general formulation in Aki and Richards (2002), who describe the discontinuities in $r_4$, $r_3$ and $l_2$ by means of the functions $f_R(k, m)$, $f_S(k, m)$ and $f_T(k, m)$. The particular cases of point forces acting in directions $z$ and $x$ are described by $f_R^m(k, m = 0) = F_z$, $f_S(k, m = 1) = -F_x/2 = -f_S(k, m = -1)$, and $f_T(k, m = 1) = iF_x/2 = f_T(k, m = -1)$. These functions vanish for any other value of $m$.

Therefore, to obtain the body-wave terms of $G_{33}$, a linear combination of the vectors $(Y_1^{PSV})_{j1}$ and $(Y_1^{PSV})_{j2}$ fulfilling the system of equations $(r_1, r_2, 0, \delta r_4)^T = Y_1^{PSV}(a, b)^T$ has to be found. The coefficients $(a, b)$ can be worked out from the third and fourth equations and used to write the factor $r_2(\omega, k)$ of the integrand as

$$r_2(\omega, k) = \frac{(Y_1^{PSV})_{31}(Y_1^{PSV})_{22} - (Y_1^{PSV})_{32}(Y_1^{PSV})_{21}}{(Y_1^{PSV})_{31}(Y_1^{PSV})_{42} - (Y_1^{PSV})_{41}(Y_1^{PSV})_{32}}, \tag{24}$$

where $F_z$ has been replaced with 1. To obtain the contributions of P and SV waves to the radial motion, $r_1(\omega, k)$ is worked out from $(r_1, r_2, \delta r_3, 0)^T = Y_1^{PSV}(a, b)^T$. Proceeding as in the previous case, it results:

$$r_1(\omega, k) = \frac{(Y_1^{PSV})_{42}(Y_1^{PSV})_{11} - (Y_1^{PSV})_{41}(Y_1^{PSV})_{12}}{(Y_1^{PSV})_{31}(Y_1^{PSV})_{42} - (Y_1^{PSV})_{41}(Y_1^{PSV})_{32}}. \tag{25}$$

Finally, in the case of SH waves, the system of equations to be solved is $(l_1, \delta l_2)^T = Y_1^{SH} a$, with $\delta l_2 = \delta \sigma_{zx} = F_x = 1$. The solution is:

$$l_1 = a \left(Y_1^{SH}\right)_1 = \delta l_2 \left(Y_1^{SH}\right)_1 / \left(Y_1^{SH}\right)_2 = \left(Y_1^{SH}\right)_1 / \left(Y_1^{SH}\right)_2 \qquad (26)$$

To illustrate the computations of surface-waves and body-waves parts of $\text{Im}[G_{jj}(0;0;\omega)]$, we show the individual contributions of each type of wave for several shallow models listed in Table 1. Models 2 and 3 (called M5 and M6 in Pei, 2007) consist of three thin layers (35 m total) overlying a thicker fourth layer (95 m). These two models present low and high velocity zones in the second layer, respectively, and the velocity contrasts are moderate.

As shown in Figure 5, the horizontal component is dominated by surface waves from 0.7 Hz in both models, with Love waves being the main component. SH waves are the main contribution at lower frequencies. For Model 3, the fundamental Love mode dominates the surface wavefield, whereas the first higher mode becomes the dominant Love mode at 14.5 Hz for Model 2, with Rayleigh waves being the main contribution in a narrow transition band around this frequency. The vertical component is dominated by Rayleigh waves of the fundamental mode in both models. As shown in Figs. 5e-f, the surface waves control the overall shape of the HVSRN except for frequencies below the fundamental SH resonance of the model. Resonances of body waves also introduce some bumps in the H/V ratio, such as the visible SH resonance at 4.9 Hz for Model 2. In spite of the smaller effects on the overall shape of the spectral ratio, the computation of body wave contributions is often the most time-consuming part of the forward problem. For example, between 5000 and 10000 evaluations of the integrand were carried out to obtain a perfect estimation of SH and P-SV terms for Model 3 in the frequency ranges shown in Fig. 5. Those integrals consumed the 91 – 97% of the total computing time. In inverse problem, the user may assess the appropriateness of neglecting body wave contributions to speed up repetitive forward calculations on the basis of an *a priori* estimation of the ground model and considering the size of the experimental uncertainties.

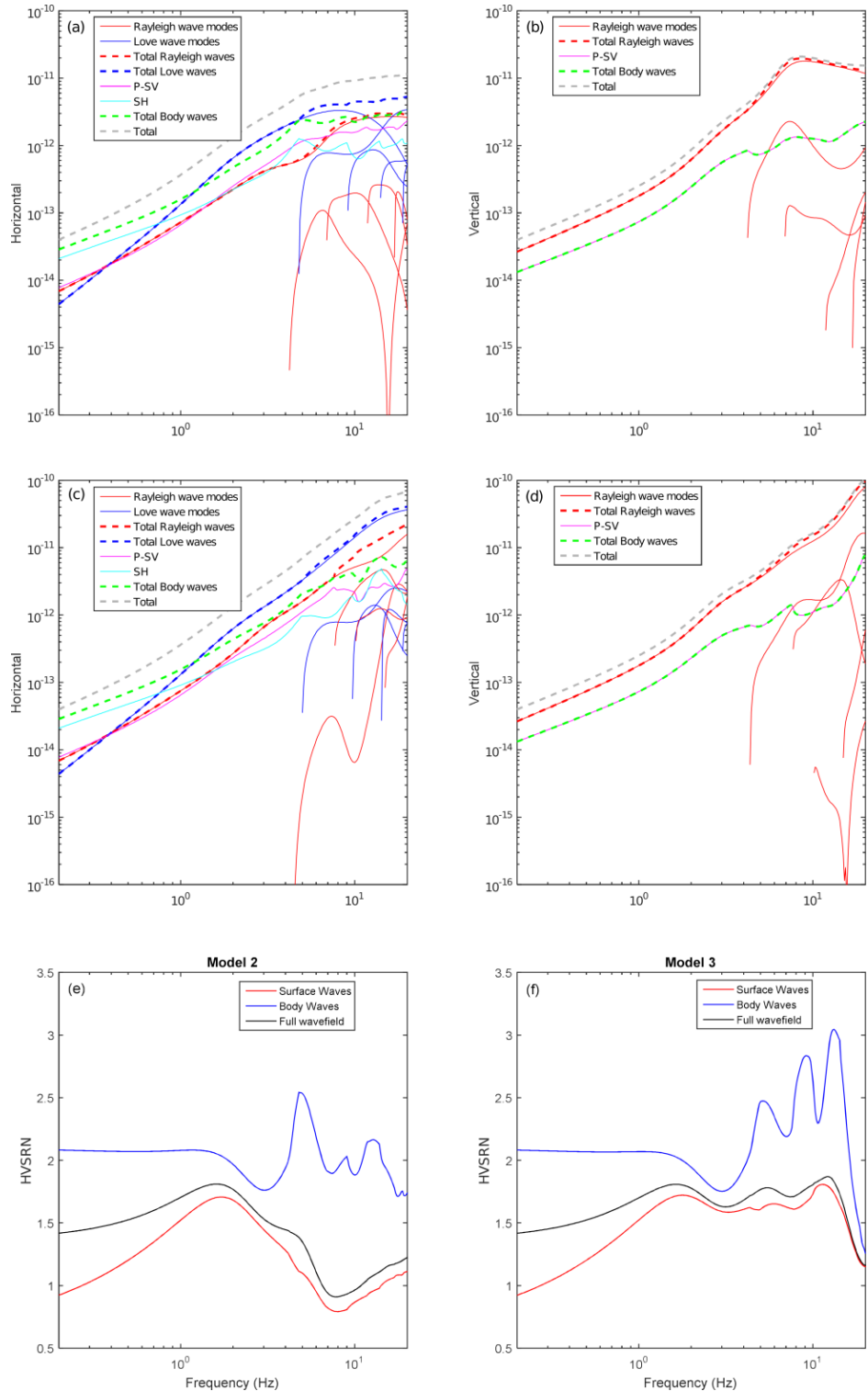

**Figure 5.** Contribution of different waves to the imaginary parts of $G_{11}$ and $G_{33}$ for coinciding source and receiver located at surface. Panels (a) and (b) correspond to horizontal and vertical components for Model 2 (Table 1). Panels (c) and (d) correspond to Model 3. The HVSRN are shown in panels (e) and (f) for models 2 and 3, using surface waves only, body waves only and full wavefield.

It should be noted that the models consider here are purely elastic. However, it is known that pronounced viscoelasticity, which is often found in shallow structures, may considerably change the properties of seismic waves. Our experiments based on simple models with high impedance contrasts (e.g. Sanchez-Sesma et al., 2011; Salinas et al., 2014; Sanchez-Sesma, 2016) show that the main effect of the anelastic attenuation is the decreasing of the main peak amplitude as the quality factor decreases. For example, Sanchez-Sesma (2016) has reported a decrement of 23% in amplitude at the peak for a quality factor of 200. Contrary to some other approaches (Herak, 2008), the effects of attenuation were almost negligible at higher frequencies, at which the Green's functions are dominated by surface waves, and the effects on the overall shape remain moderate.

## 3. Inversion of HVSRN and of surface wave velocities

In a general sense, the inversion of geophysical data consists in finding properties of the so called *a posteriori* probability distribution, which measures the probability of any model **m** (a ground profile) of being the true model on the basis of its ability to fit a set of observations $d_j^{obs} \pm \sigma_j$ and fulfilling *a priori* constraints. The method for forward calculation, $d_j^{theo}(\mathbf{m})$, is assumed to be known. Most of the applications are intended to characterize this probability distribution by means of the maximum likelihood model, the mean (expected) model and/or the covariance matrix for the model parameters.

The inversion software provided here has been written in Matlab® and includes implementations of three main inversion algorithms: i) Monte Carlo sampling ii) the Simulated Annealing method (SA) and iii) the Interior Point method (IP), together with a suitable graphic user interface. In addition, two simpler methods such as the Random Search (RS) and the Downhill-Simplex are also available. These methods can be sequentially applied following the program flow chart sketched in Figure 6.

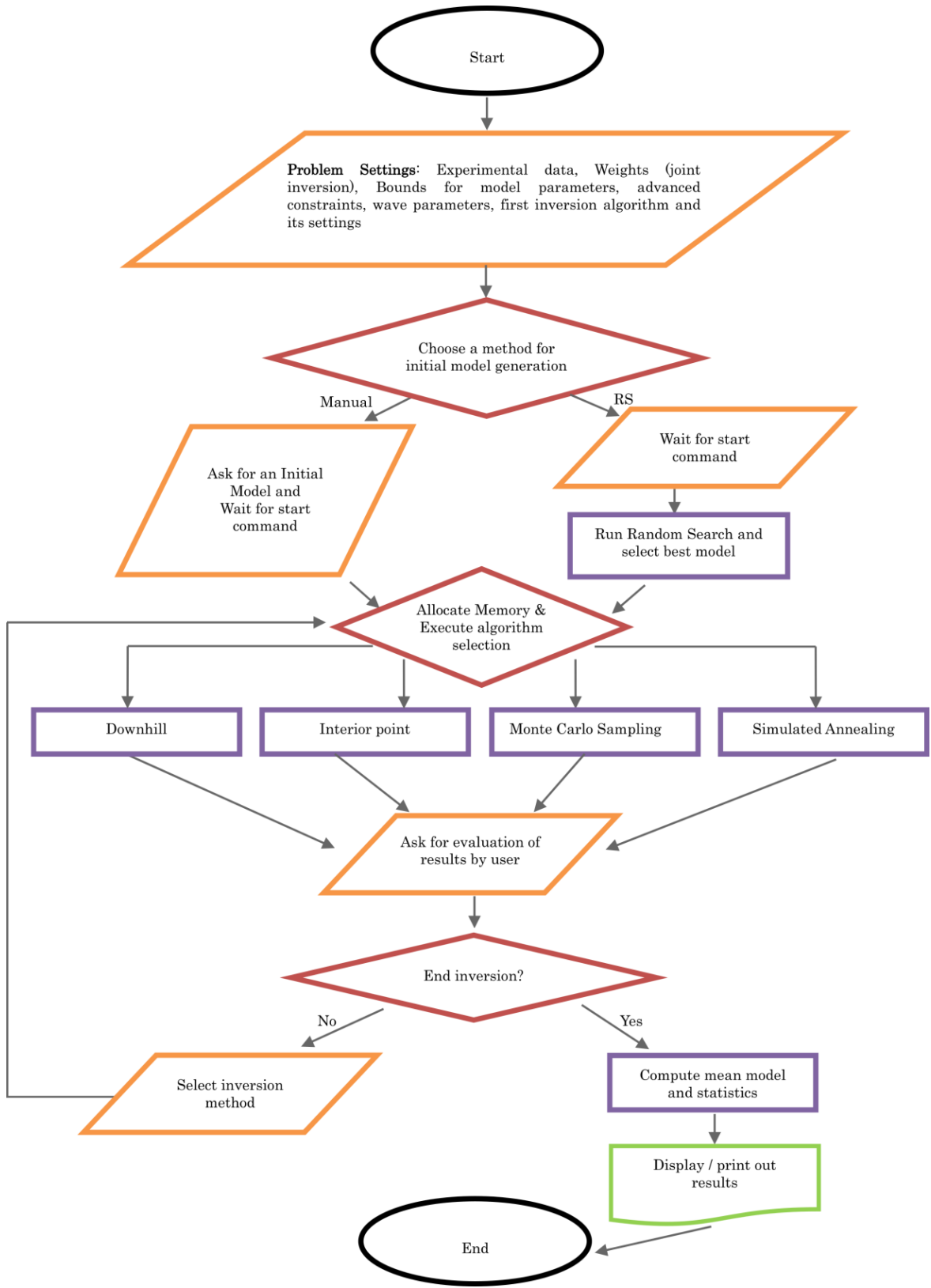

**Figure 6.** Flow chart of the inversion procedure.

The interface allows both independent and joint inversion of HVSRN and Rayleigh- or Love-wave phase or/and group velocities. Since only variations in the elastodynamic properties with depth are considered, our procedure requires that the structure beneath the experimental setting may be locally approximated by a horizontally layered medium. If the dataset includes surface wave velocities and HVSRNs, the similarity among the spectral ratios for all the stations of the array used for dispersion curve retrieval can be stated as a consistency control. The main features of these implementations are described in this section.

### 3.1 Random Search

A random search on the model space can be performed alone or as a first stage of a more complicated inversion procedure. The user has to set up suitable intervals for the ground parameters ($\alpha_j$, $\beta_j$, $\rho_j$, $h_j$ and Poisson's ratios) as *a priori* information. Uniform probability is assumed for layer thicknesses and densities within the stated ranges as well as for either $\alpha_i$ or $\beta_i$, depending on the program settings. The remaining velocity type is also uniformly sampled even though the actual range could be made narrower to be compatible with the Poisson's ratio intervals. In addition, it is possible to impose downward-increasing $\alpha(z)$ or $\beta(z)$ velocity profiles. Instead of rejecting invalid profiles among the random models, the semi-analytical approach described in Appendix A has been followed to implement these constraints.

After a number of trials defined by the user, the model presenting the lowest value of the misfit, $E(\mathbf{m})$, defined as

$$E(\mathbf{m}) = \sum_j \left(d_j^{obs} - d_j^{theo}(\mathbf{m})\right)^2 / \sigma_j^2 \tag{27}$$

is selected. For joint inversion of an HVSRN together with a dispersion curve, equation (27) is slightly generalized to

$$E(\mathbf{m}) = 2(1-w_c) \sum_{j \in n_{HV}} \left(HV_j^{obs} - HV_j^{theo}(\mathbf{m})\right)^2 / \sigma_{HVj}^2 + 2w_c \sum_{j \in n_c} \left(c_j^{obs} - c_j^{theo}(\mathbf{m})\right)^2 / \sigma_{cj}^2, \quad (28)$$

where $n_{HV}$ and $n_c$ are the respective number of samples and an adjustable weight parameter $w_c$ ranging from 0 to 1 controls the relative weight of the dispersion curve in the global misfit. The default choice of $w_c = 0.5$ guarantees the same weight for all the samples, regardless of the observable they represent. On the contrary, $w_c$ can be used to equalize the sensitivity of $E(\mathbf{m})$ to spectral ratios and velocities, which could be retrieved for a very different number of samples. To do so, the software provides the option of making an inverse weighting fulfilling $w_{HV}/w_c = n_c/n_{HV}$, where $w_{HV}$ represents $(1-w_c)$. In particular, it takes

$$w_c = \frac{n_{HV}}{n_{HV} + n_c} = \frac{n_{HV} n_c}{n_{HV} + n_c} \frac{1}{n_c}, \text{ so that } w_{HV} = \frac{n_{HV} n_c}{n_{HV} + n_c} \frac{1}{n_{HV}}. \text{ In the case of coinciding } n_{HV}$$

and $n_c$ then $w_c$ and $w_{HV}$ equal 0.5 and the unweighted expression (27) is recovered.

### 3.2 Simulated Annealing and Monte Carlo Sampling

The SA inversion method (Kirkpatrick et al., 1983) is inspired in the process of heating a solid until it melts and then cooling very slowly until the substance reaches the state of lowest energy, forming a perfect crystal. This technique uses the Metropolis algorithm (Metropolis et al., 1953; Hastings, 1970) to sample the model space $M$ with a Gibbs-type probability density function

$$G(\mathbf{m}, T) = \exp\left(\frac{-E(\mathbf{m})}{T}\right) \bigg/ \int_M \exp\left(\frac{-E(\mathbf{m})}{T}\right) d\mathbf{m}. \quad (29)$$

In equation (29), $E(\mathbf{m})$ resembles the energy, and $T$ a fixed value which is decreased slowly and corresponds to the temperature in the thermodynamic analog (the Boltzmann constant is taken equal to one).

The cooling schedule, defined by the number of temperatures, its initial value and reduction ratio as well as the number of model per temperature (Markov's chain length) are set by the user. In addition to the best fitting model, the software computes the mean model $\overline{\mathbf{m}}$, its uncertainties and the normalized covariance matrix **c** for a set of iterations performed at the lower temperature reached. The relevant formulae are:

$$\overline{\mathbf{m}} = \int_M \mathbf{m} \, G(\mathbf{m}) \, d\mathbf{m} \tag{30}$$

and

$$c_{ij} = \left|C_{ij}\right| / \sqrt{C_{ii} C_{jj}}, \quad \mathbf{C} = \int_M (\mathbf{m} - \overline{\mathbf{m}})(\mathbf{m} - \overline{\mathbf{m}})^T G(\mathbf{m}) \, d\mathbf{m}, \tag{31}$$

where **m** is written as a column vector. Since the model space is sampled according to the distribution $G(\mathbf{m})$, the integrals turn into the unweighted averages of **m** and $(\mathbf{m} - \overline{\mathbf{m}})(\mathbf{m} - \overline{\mathbf{m}})^T$ respectively, over the set of models obtained in this way.

This method has been successfully used for inversion of surface-waves dispersion curves by Iglesias (2000) and Pei et al. (2007) among others. Comparisons between the SA and two more heuristic methods as well as with linearized inversion have been carried out by Yamanaka (2005) and Pei (2007), respectively. Yamanaka`s numerical experiments showed that SA provides a rapid convergence and finds models with the smallest misfits.

With the above definition of $E(\mathbf{m})$, the *a posteriori* probability density in the case of independent Gaussian uncertainties is proportional to $P(\mathbf{m}, T = 2)$. The exploration at this constant temperature is implemented as a separate option of the software and referred to as Random Monte Carlo sampling (e.g. Mosegaard and Tarantola, 1995). In this case, $\overline{\mathbf{m}}$ and $c_{ij}$ calculated from Eqs. (30) and (31) have a clear statistical meaning. Note that the introduction of the decreasing temperature in the SA method can be thought as a way of distorting the *a posteriori* probability distribution until it sharply peaks at the maximum likelihood (minimum energy) model (Tarantola, 2005). In this way, the neighborhood of this model is virtually the only sampled region at the end of the cooling process.

**3.3 Local optimization methods**

The built-in implementation of the interior-point method for non-linear constrained optimization (e.g. Waltz et al., 2006) usually converges faster than the SA algorithm (and much faster than the RS) to a minimum of $E(\mathbf{m})$. Since it is prone to converge to local minima, this algorithm would be suitable when the a priori information allows for significant restrictions of the model space (encoded by the user) or when the solution is known to be well approximated by the initial model. This method also performs well for joint inversion of HVSR and dispersion curves, due to the implied reduction of the set of equivalent solutions (see Piña-Flores et al., 2016 for further discussion on non-uniqueness).

The procedure is based on the Matlab implementation of this method through *fmincon* command, which attempts to find the minimum of constrained nonlinear multivariable functions. The user has to set up the termination criterion in terms of minimum meaningful variations in model parameters and misfit decrements. The maximum number of iterations and misfit evaluations can also be limited. The bounds imposed to the model parameters as well the optional constraints related with increasing $\alpha(z)$ or $\beta(z)$ functions are coded by means of the inequalities which define the constraints in this optimization method.

A basic implementation of the downhill simplex method (Nelder and Mead, 1965) based on a built-in Matlab procedure has also been incorporated to the interface. Even though this derivative-free algorithm is aimed at unconstrained minimization problems, the ground model constraints are taken into account through misfit penalization of models for which they are violated. The speed of this method is very good for low and moderate model space dimensionality.

## 4. Test with real data

The reliability of H/V ratios obtained under the diffuse field assumption has been already stated by Sánchez-Sesma et al. (2011) and Salinas et al. (2014) among others. These two studies deal with H/V curves of relatively simple shape associated with resonances of a shallow soft layer (as soft as 70 m/s for $\beta_1$). In this section, the outputs of the presented software and its suitability are illustrated for a sample HVSRN measured in a very different environment.

The measurements used here were performed in the context of seismic investigations in Campo de Dalías, a large coastal plain in SW Spain and correspond to a test site located at El Ejido town. Previous active multichannel seismic surveys and borehole analysis revealed the stratigraphic record in the vicinity of the measurement point. The shallow layers are formed by anthropogenic fillings of insignificant thickness overlying Pliocene materials (calcarenites, silts and marly limestones). The Pliocene-aged layer is underlain by the thickest Neogene unit, of Messinian age (Pedrera et al., 2015), and composed at its base of lower Messinian silts and marls overlain by gypsum layers, marly limestones and silts. An older and thinner layer made up of late Tortonian calcarenites and conglomerates rests on the basement, which is attributed to the Permo-Triassic metamorphic rocks of the Alpujárride Complex (Betic Cordillera).

We used a CMG-6TD broad-band seismometer. The recording time was of 30 minutes and the sampling rate was 100 sps. Traces were first divided in a set of overlapping windows of 40.48s. Then, the HVSRN was calculated following the procedure stated by Sánchez-Sesma et al. (2011), which involves normalization of each time window by its total energy, calculation of the power spectral density for each component and the subsequent computation of the spectral ratio according with Eq. (1). The H/V curve (black line in Fig. 7a) shows a somewhat complicated shape with a clear main peak at 0.65-0.7 Hz and a broad secondary bump between 1.3 and 8.0 Hz, approximately. In turn, at least a pair of minor peaks can be identified in that latter band.

The inversion has been performed combining SA with the local methods. The allowed ranges for the ground parameters $\alpha$ and $\rho$ were based on previous work performed by Marín-Lechado (2005) in Campo de Dalías. Once a good fitting of the data was obtained, 4000 more models were generated by Monte Carlo sampling to compute the mean model, the standard deviation and the covariance matrix (Fig 7). The obtained model is consistent with the information available from previous studies. For example, the two-way P-wave travel time down to the bottom of the layer interpreted as Messinian marls (418 ms) agrees very well with the corresponding value derived from active-source seismic methods (Plata et al., 2004, p. 164), and the depth of the basement matches the results obtained by González et al. (2003).

## 5. Concluding remarks

An algorithm for forward calculation of HVSRN based on the diffuse field approximation (Sánchez-Sesma et al., 2011) has been described and implemented in a FORTRAN 90 code. This method provides a full-wavefield interpretation of this observable and it is compatible with the recent developments in ambient noise interferometry.

An orthonormalization technique has been used to preserve the stability of the code at high frequency, preventing from underflow/overflow problems in the susceptible steps of the algorithm: computation of dispersion curves, medium responses to surface-wave modes and contribution of body waves.

The separate calculation of each contribution to the wavefield allows a better understanding of the physics. Moreover, it would make feasible to state corrections in cases of deviations of the wavefield composition from the energy ratios stated by the equipartition principle. Examples of this behavior have been reported by Nakahara and Margerin (2011) among others.

This code has been incorporated into a program for inversion of HVSRN which supports joint inversion of the spectral ratio and the surface waves dispersion curves. This approach has been demonstrated to reduce the tradeoff between thicknesses and velocities inherent to inversion of H/V ratios and improves the sensitivity to velocity contrasts of profiles obtained from dispersion curves, and in particular to the basement velocity (e.g. Arai and Tokimatsu, 2005; Parolai et al., 2005; Piña-Flores et al., 2016). The program allows for applying Monte Carlo sampling of the model space, the SA algorithm as well as local methods. The numerical calculation of the body waves integrals is often the most time-consuming part of the forward problem, even though these terms often have minor effects on the HVSRN curves. In very intensive problems, the software can be used for doing a preliminary inversion based on surface-waves components, letting final refinements with full wavefield computations for a second stage.

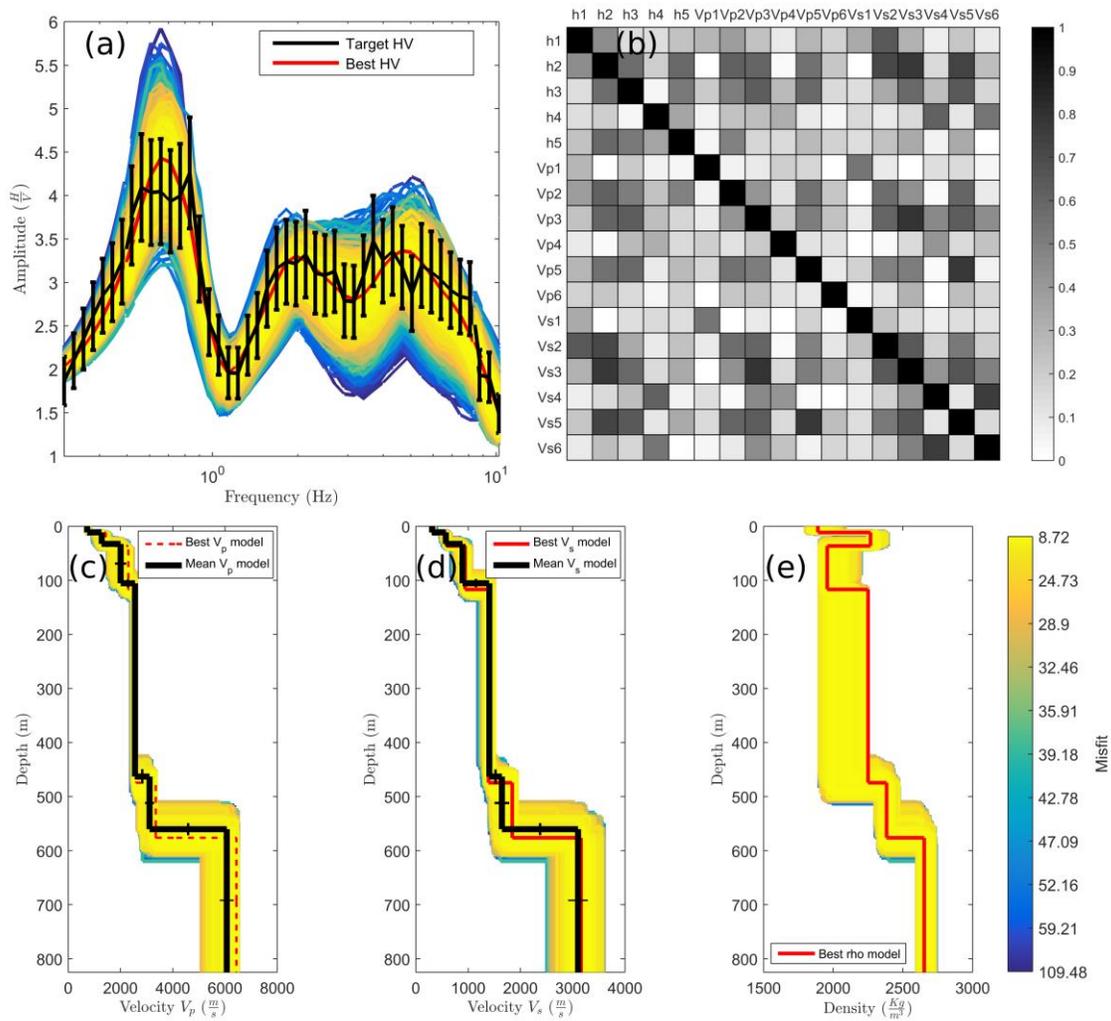

**Figure 7.** Example of HVSRN inversion for a test site (Campo de Dalías, SW Spain). Black line and black segments in panel (a) represent the experimental spectral ratio and its standard deviation. The colored curves in (a) show the forward calculations for the corresponding sampled models shown in panels (c) to (e), which were evaluated during a final application of the Monte Carlo Sampling method. Red lines show the best fitting model (panels c to e) and the corresponding theoretical HVSRN (panel a). The mean model **m** and the standard deviations of its parameters ($C_{jj}$) are shown in panels c-d with black lines, whereas panel (b) shows the normalized covariance matrix **c**.

The application of this software to real data obtained at a moderately deep sedimentary structure provided good fitting of the experimental curve, including secondary peaks and features. The results were consistent with previous information obtained from boreholes and conventional seismic methods.

## Acknowledgements

This research has been partially supported by the Spanish Ministry of Economy and Competitiveness under grants CGL2014-59908 and CGL2010-16250, the European Union with FEDER and the DGAPA-UNAM Project IN104712. Partial support is acknowledged from the AXA Research Fund. The scheme for root bracketing has been adapted from the freeware Geopsy codes.

## References

Aki, K., 1957. Space and time spectra of stationary stochastic waves, with special reference to microtremors. Bulletin of the Earthquake Research Institute 35, 415-456.

Aki, K., Richards, P.G., 2002. Quantitative Seismology, 2nd edn. University Science Books, Susalito, CA, 700 pp.

Arai, H., Tokimatsu, K., 2004. S-wave velocity profiling by inversion of microtremor H/V spectrum. Bulletin of the Seismological Society of America 94, 53-63.

Arai, H., Tokimatsu, K., 2005. S-wave velocity profiling by joint inversion of microtremor dispersión curve and horizontal-to-vertical (H/V) spectrum. Bulletin of the Seismological Society of America 95, 1766-1778.

Ben-Menahem, A., Singh, S.J., 2000. Seismic Waves and Sources, 2nd edn. Dover, New York, NY, 1136 pp.

Buchen, P.W., Ben-Hador, R., 1996. Free-mode surface-wave computations. Geophysical Journal International 124, 869-887.

Denolle, M.A., Dunham, E.M., Beroza, G.C., 2012. Solving the surface-wave eigenproblem with Chebyshev spectral collocation. Bulletin of the Seismological Society of America 102(3), 1214–1223.

Dunkin, J.W., 1965. Computation of modal solutions in layered, elastic media at high frequencies. Bulletin of the Seismological Society of America 55, 335-358.

Fäh, D., Kind, F., Giardini, D., 2003. Inversion of local S-wave velocity structures from average H/V ratios and their use for the estimation of site-effects. Journal of Seismology 7, 449-467.

García-Jerez, A., Luzón, F., Sánchez-Sesma, F.J., Lunedei, E., Albarello, D., Santoyo, M.A., Almendros, J., 2013. Diffuse elastic wavefield within a simple crustal model. Some consequences for low and high frequencies. Journal of Geophysical Research 118, 5577–5595.

García-Jerez, A., Sánchez-Sesma, F.J., 2015. Slowly-attenuating P-SV leaky waves in a layered elastic halfspace. Effects on the coherences of diffuse wavefields. Wave Motion 54, 43–57.

González, A., Domínguez, P., Franqueza, P., 2003. Sistema costero de Sierra de Gádor. Observaciones sobre su funcionamiento y relaciones con los ríos Adra y Andarax, y con el mar. In: López, J.A., De la Orden, J.A., Gómez, J.D., Ramos, G., Mejías, M., Rodríguez, L. (Eds.) Tecnología de la Intrusión de Agua de Mar en Acuíferos Costeros: Países Mediterráneos, IGME, Madrid, pp. 423-432.

Harkrider, D.G., 1964. Surface waves in multilayered elastic media. Part 1. Bulletin of the Seismological Society of America 54, 627-679.

Harkrider, D.G., Anderson, D.L., 1966. Surface wave energy from point sources in plane layered earth models. Journal of Geophysical Research 71, 2967-2980.

Herak, M., 2008. ModelHVSR - A Matlab® tool to model horizontal-to-vertical spectral ratio of ambient noise. Computers & Geosciences 34, 1514–1526.

Herrmann, R.B., Ammon, C.J., 2003. Computer programs in seismology. An overview of synthetic seismogram computation. St. Louis Univ., St. Louis, Mo., 183 pp.

Iglesias, A., 2000. Aplicación de algoritmos genéticos y Simulated Annealing para invertir la dispersión de ondas superficiales: modelos promedio de la corteza terrestre en el sur de México. Unpublished BSc. Thesis, Universidad Nacional Autónoma de México, México.

Kanai, K., Tanaka, T., Osada, K., 1954. Measurement of the microtremor I. Bulletin of the Earthquake Research Institute of Tokyo University 32, 199–210.

Kawase, H., Matsushima, S., Satoh, T., Sánchez-Sesma, F.J., 2015. Applicability of theoretical horizontal-to-vertical ratio of microtremors based on the diffuse field concept to previously observed data. Bulletin of the Seismological Society of America 105(6), 3092-3103.

Kirkpatrick, S., Gelatt Jr., C.D., Vecchi, M.P., 1983. Optimization by simulated annealing. Science 220, 671-680.

Knopoff, L., 1964. A matrix method for elastic wave problems. Bulletin of the Seismological Society of America 54, 431-438.

Lachet, C., Bard, P.Y., 1994. Numerical and theoretical investigations on the possibilities and limitations of Nakamura's technique. Journal of Physics of the Earth 42, 377397.

Lontsi, A.M., Sánchez-Sesma, F.J., Molina-Villegas, J.C., Ohrnberger, M., Krüger, F., 2015. Full microtremor H/V($z$, $f$) inversion for shallow subsurface characterization. Geophysical Journal International 202(1), 298-312.

Lunedei, E., Albarello, D., 2009. On the seismic noise wavefield in a weakly dissipative layered Earth. Geophysical Journal International 177, 1001–1014.

Lunedei, E., Malischewsky, P., 2015. A review and some new issues on the theory of the H/V technique for ambient vibrations, in: Ansal, A. (Ed.), Perspectives on European Earthquake Engineering and Seismology, Geotechnical, Geological and Earthquake Engineering, Vol. 39. Springer, Heidelberg, pp. 371-394. DOI 10.1007/978-3-31916964-4_15.

Margerin, L., 2009. Generalized eigenfunctions of layered elastic media and application to diffuse fields. Journal of the Acoustical Society of America 125, 164–174.

Marín-Lechado, C., 2005. Estructura y evolución tectónica reciente del Campo de Dalías y de Níjar en el contexto del límite meridional de las Cordilleras Béticas orientales. Ph.D. Dissertation, Universidad de Granada, Granada, Spain, 266 pp. In Spanish.

Mosegaard, K., Tarantola, A., 1995. Monte Carlo sampling of solutions to inverse problems. Journal of Geophysical Research 100B7, 12431–12447.

Nakahara, H., Margerin, L., 2011. Testing equipartition for S-wave coda using borehole records of local earthquakes. Bulletin of the Seismological Society of America 10, 2243–2251.

Nakamura, Y., 1989. A method for dynamic characteristics estimation of subsurface using microtremors on the ground surface. Quarterly Report of Railway Technical Research Institute (RTRI) 30, 25-33.

Nelder, J.A., Mead, R., 1965. A Simplex Method for Function Minimization. Computer Journal 7, 308–313.

Parolai, S., Picozzi, M., Richwalski S.M., Milkereit, C., 2005. Joint inversion of phase velocity dispersion and H/V ratio curves from seismic noise recordings using a genetic algorithm, considering higher modes. Geophysical Research Letters 32, L01303, doi:10.1029/2004GL021115.

Pedrera, A., Marín-Lechado, C., Galindo-Zaldívar, J., Lobo, F.J., 2015. Smooth folds favoring gypsum precipitation in the Messinian Poniente marginal basin (Western Mediterranean). Tectonophysics, In press. doi:10.1016/j.tecto.2015.05.019

Pei, D., 2007. Modeling and inversion of dispersion curves of surface waves in shallow site investigations. PhD Dissertation, University of Nevada, Reno, 165 pp.

Pei, D., Louie, J.N., Pullammanappallil, S.K., 2007. Application of simulated annealing inversion on high-frequency fundamental-mode Rayleigh wave dispersion curves. Geophysics 72(5), R77–R85.

Piña-Flores, J., Perton, M., García-Jerez, A., Carmona, E., Luzón, F., Molina-Villegas, J.C., Sánchez-Sesma, F.J., 2016. The inversion of spectral ratio *H/V* in a layered system using the Diffuse Field Assumption (DFA). Geophysical Journal International. Submitted.

Plata-Torres, J.L., Lopez-Sopeña, F., Aracil-Avila, E., Lopez-Mendieta, F.J., 2004. Estudio estructural del acuífero profundo del Campo Dalías (Almería) mediante sísmica de reflexión - Tomo I. Instituto Geológico y Minero de España, Madrid, Spain, 221 pp. URL: http://info.igme.es/ConsultaSID/presentacion.asp?Id=106040

Poggi, V., Fäh, D., Burjanek, J., Giardini, D., 2012. The use of Rayleigh-wave ellipticity for site-specific hazard assessment and microzonation: application to the city of Lucerne, Switzerland. Geophysical Journal International 188(3), 1154-1172.

Rivet, D., Campillo, M., Sánchez-Sesma, F.J., Shapiro, N.M., Singh, S.K., 2015. Identification of surface wave higher modes using a methodology based on seismic noise and coda waves. Geophysical Journal International 203(2), 856-868.

Salinas, V., Luzón, F., García-Jerez, A., Sánchez-Sesma, F. J., Kawase, H., Matsushima, S., Suarez, M., Cuellar, A., Campillo, M., 2014. Using diffuse field theory to interpret the H/V spectral ratio from earthquake records in Cibeles seismic station, Mexico City. Bulletin of the Seismological Society of America 104(2), 995–1001.

Sánchez-Sesma, F.J., 2016. Modeling and inversion of the microtremor H/V spectral ratio: physical basis behind the diffuse-field approach. In: Proceedings 5th IASPEI / IAEE International Symposium: Effects of Surface Geology on Seismic Motion, Taipei, Taiwan, p. 1-8.

Sánchez-Sesma, F.J., Campillo, M., 2006. Retrieval of the Green function from crosscorrelation: The canonical elastic problem. Bulletin of the Seismological Society of America 96, 1182-1191.

Sánchez-Sesma, F.J., Pérez-Ruiz, J.A., Luzón, F., Campillo, M., Rodríguez-Castellanos, A., 2008. Diffuse fields in dynamic elasticity. Wave Motion 45, 641–654.

Sánchez-Sesma, F.J., Rodríguez, M., Iturrarán-Viveros, U., Luzón, F., Campillo, M., Margerin, L., García-Jerez, A., Suarez, M., Santoyo, M.A., Rodríguez-Castellanos, A., 2011. A theory for microtremor H/V spectral ratio: application for a layered medium. Geophysical Journal International 186, 221–225.

Shapiro, N.M., Campillo, M., 2004. Emergence of broadband Rayleigh waves from correlations of the ambient seismic noise. Geophysical Research Letters 31, L07614, doi 10.1029/2004GL019491.

Snieder, R., 2004. Extracting the Green's function from the correlation of coda waves: A derivation based on stationary phase. Physical Review E 69, 046610.

Spica, Z., Caudron, C., Perton, M., Lecocq, T., Camelbeeck, T., Legrand, D., Piña-Flores, J., Iglesias, A., Syahbana, D.K., 2015. Velocity models and site effects at Kawah Ijen volcano and Ijen caldera (Indonesia) determined from ambient noise cross-correlations and directional energy density spectral ratios. Journal of Volcanology and Geothermal Research 302, 173-189.

Tarantola, A., 2005. Inverse Problem Theory and Methods for Model Parameter Estimation. Society of Industrial and Applied Mathematics, Philadelphia, PA, 342 pp.

Tokimatsu, K., Tamura, S., 1995. Contribution of Rayleigh and body waves to displacement induced by a vertical point force on a layered elastic half-space. Journal of Structural and Construction Engineering 476, 95-101.

Waltz, R.A., Morales, J.L., Nocedal, J., Orban, D., 2006. An interior algorithm for nonlinear optimization that combines line search and trust region steps. Mathematical Programming 107(3), 391–408.

Wang, R., 1999. A simple orthonormalization method for stable and efficient computation of Green's functions. Bulletin of the Seismological Society of America 89, 733-741.

Wapenaar, K., 2004. Retrieving the elastodynamic Green's function of an arbitrary inhomogeneous medium by cross correlation. Physical Review Letters 93, 254301.

Wathelet, M., 2005. Array recordings of ambient vibrations: surface-wave inversion. PhD Dissertation, University of Liège, Belgium, 161 pp.

Yamanaka, H., 2005. Comparison of performance of heuristic search methods for phase velocity inversion in shallow surface wave method. Journal of Environmental and Engineering Geophysics 10(2), 163-173.

**Appendix A**. Algorithm for generation of models with velocity increasing downwards.

In this software, the velocities of the layers, $v_j$, $j = 1, 2, \ldots, N$ are considered as independent variables with uniform probability densities in predefined intervals [$v_{j\min}$ $v_{j\max}$]. In mathematical terms, the probability density for the combination $(v_1, v_2, \ldots, v_N)$ is $p(v_1, v_2, \ldots, v_N) = p_1(v_1) p_2(v_2) \ldots p_N(v_N)$, where $p_j(v)$ is

$$p_j(v) = \begin{cases} (v_{j\max} - v_{j\min})^{-1}, & v_{j\min} \leq v \leq v_{j\max} \\ 0, & \text{otherwise} \end{cases}, \quad (A1)$$

which can be easily sampled with a generator of pseudo-random uniformly-distributed numbers.

To let the user introduce an *a priori* condition of increasing velocity with depth, the algorithm might simply discard those models for which that condition fails (i.e. $v_{j+1} < v_j$, for some pair of successive layers). Nevertheless, this procedure becomes very inefficient as the number of layers increases so that an alternative semi-analytical approach has been implemented.

Considering the statistics described by equations (A1-2), the probability of generating a profile with velocity of the $(j+1)$-th layer greater than $v_j$ and increasing velocities in the underlying layers can be represented by

$$I_j(\xi) = \int_\xi^{v_{j+1\_\max}} d\xi \, p_{j+1}(\xi) \ldots \int_\xi^{v_{N-1\_\max}} d\xi \, p_{N-1}(\xi) \int_\xi^{v_{N\_\max}} d\xi \, p_N(\xi) \quad (A2)$$

$I_j(\xi)$ defined in this manner is a piecewise polynomial function.

Once the expression of $I_1(\xi)$ has been calculated, the velocity of the upper layer is sampled. To do so, we use a uniform random number generator and the transformation method of probability distributions (e. g. Press et al. 2007, section 7.3.2). First, the

cumulative distribution function for $v_1$, also of piecewise-polynomial type, is calculated as

$$P(v_1 | v_1 \leq v_2 \leq ... \leq v_N) = 1 - \frac{\int_{v_1}^{v_{1\max}} d\xi\, I_1(\xi)}{\int_{v_{1\min}}^{v_{1\max}} d\xi\, I_1(\xi)}, \qquad v_{1\min} \leq v_1 \leq v_{1\max} \qquad (A3)$$

Then, for a sample $s$ of the uniform distribution in [0 1], the corresponding sample $v_{1s}$ of the target distribution is calculated as $v_{1s} = P^{-1}(s | v_1 \leq v_2 \leq ... \leq v_N)$. This equation is solved numerically, without an explicit calculation of $P^{-1}$.

This procedure is iterated for deeper layers down to the halfspace, taking the sample obtained for the previous layer as a lower bound for the velocity of the current one. In this way, for the $j$-th layer, equation (A3) is replaced with

$$P(v_j | v_{j-1}; v_j \leq v_{j+1} \leq ... \leq v_N) = 1 - \frac{\int_{v_j}^{v_{j\max}} d\xi\, I_j(\xi)}{\int_{v_{j-1}}^{v_{j\max}} d\xi\, I_j(\xi)}, \qquad v_{j-1} \leq v_j \leq v_{j\max}. \qquad (A4)$$

An example of this procedure is presented in Figure A1. As shown, the statistics of the models obtained in this way behave as those of the reference population, i.e. the set obtained assuming uniform distributions of velocities and subsequently discarding those models for which $v(z)$ is not a monotonically increasing function. For reference, and assuming that the lower and upper velocity limits are same for all the layers, the procedure described here is faster than the reference method for models with more than 5-6 layers (Fig. A2). The reference method becomes unfeasible for about 8 layers, when the time required for generating an acceptable model is of the order of the duration of the forward calculations.

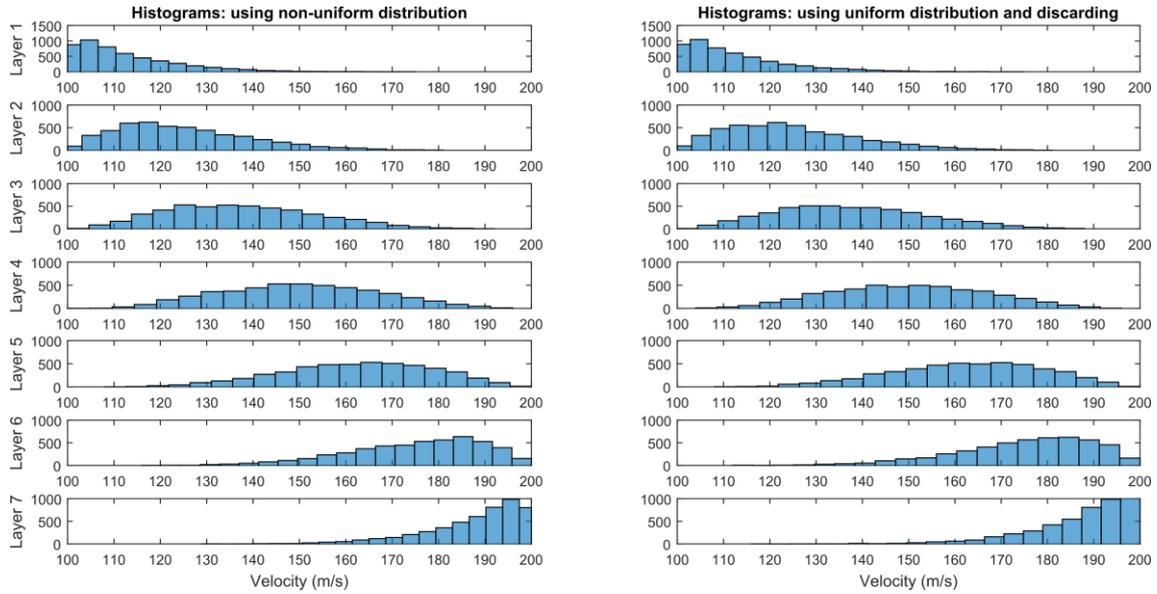

**Figure A1.** Histograms for the velocities of 5000 models consisting of six layers overlying a halfspace for which the velocity increases as depth increases. Left side panels are generated on the basis of the expressions shown in this appendix. Right side panels represent the reference set of models, with uniform distribution of the velocities within the limits of the layer and subsequent discard of invalid models. For the latter method, only the 0.02 % of the models was acceptable.

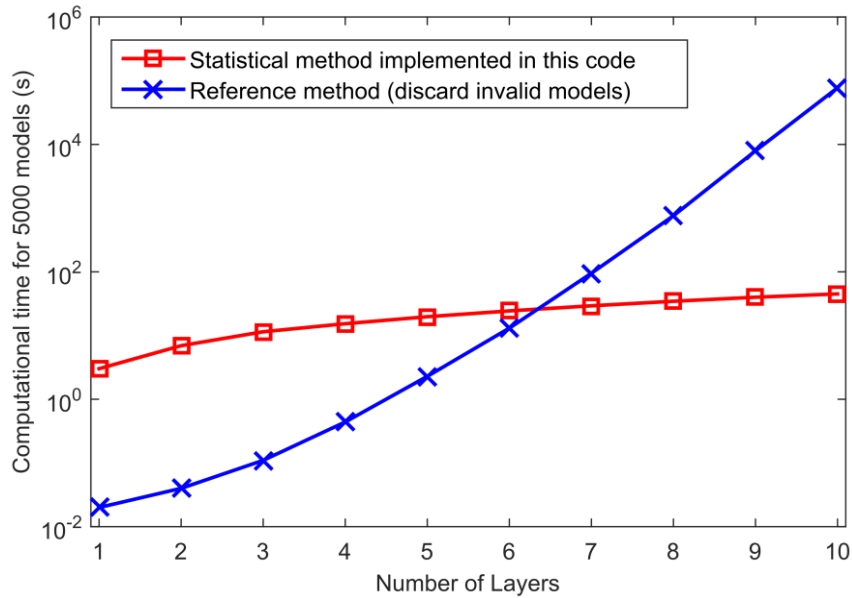

**Figure A2.** Computing time for generation of 5000 random models with monotonically increasing $v(z)$, following the two methods described here. The displayed durations are approximate values for an Intel® i7-4510U 2.0GHz processor using Matlab® code.

If the only constraint required by the user consists in assigning the higher velocity to the halfspace, we compute the piecewise polynomial function

$$P(v_N \mid v_1 \le v_N, ..., v_{N-1} \le v_N) = \frac{\int_{v_{N\min}}^{v_N} d\xi\, J(\xi)}{\int_{v_{N\min}}^{v_{N\max}} d\xi\, J(\xi)}, \qquad v_{N\min} \le v_N \le v_{N\max} \tag{A5}$$

instead, where $J(\xi) = \prod_{j=1}^{N-1} \int_{v_{j\min}}^{\xi} dv\, p_j(v)$ measures the probability of having the velocities of the upper $N$-1 layers smaller than an arbitrary value $\xi$. Once the velocity of the halfspace $v_{Ns}$ is sampled according to the distribution in (A5), the rest of the velocities are generated with uniform distributions in $[v_{j\min}\ \ v_{Ns}]$.

**References**

Press, W.P., Teukolsky, S.A., Vetterling, W.T., Flannery, B.P., 2007. Numerical Recipes: The Art of Scientific Computing, 3rd edn. Cambridge University Press, New York, NY, 1256 pp.